\documentclass[aps,pra,showpacs,twocolumn,superscriptaddress]{revtex4-2}
\usepackage{amsmath}
\usepackage{amssymb}
\usepackage{graphicx}
\usepackage{pdfsync}
\usepackage{lipsum}
\usepackage{bbm}

\usepackage{siunitx}
\usepackage{epstopdf}
\usepackage{times,txfonts}
\usepackage{easybmat}
\usepackage{mathtools}
\usepackage{upgreek}
\usepackage{appendix}
\newcommand{\ket}[1]{|#1\rangle}

\usepackage[bookmarks=false]{hyperref}
\hypersetup{colorlinks=true,citecolor=blue,linkcolor=blue,urlcolor=blue,pdfstartview=FitH,bookmarksopen=true}
\usepackage{color,soul}
\usepackage{orcidlink}

\begin{document}
\title{High-frequency dual-channel lock-in detection via rapidly oscillating driving}

\author{Kangze Li\orcidlink{0000-0001-5258-1167}}
\affiliation{College of Physics and Optoelectronic Engineering and Shanxi Key Laboratory of Precision Measurement Physics, Taiyuan University of Technology, Taiyuan 030024, China}
\affiliation{School of Mathematical Sciences and Centre for the Mathematics and Theoretical Physics of Quantum Non-Equilibrium Systems, University of Nottingham, University Park, Nottingham NG7 2RD, United Kingdom}

\author{Xu Zhao}
\affiliation{College of Physics and Optoelectronic Engineering and Shanxi Key Laboratory of Precision Measurement Physics, Taiyuan University of Technology, Taiyuan 030024, China}

\author{Liantuan Xiao\orcidlink{0000-0002-0104-8634}}
\email{xlt@sxu.edu.cn}

\affiliation{College of Physics and Optoelectronic Engineering and Shanxi Key Laboratory of Precision Measurement Physics, Taiyuan University of Technology, Taiyuan 030024, China}

\affiliation{Key Laboratory of Advanced Transducers and Intelligent Control System, Taiyuan University of Technology and Ministry of Education, Taiyuan 030024, China}

\affiliation{State Key Laboratory of Quantum Optics Technologies and Devices, Shanxi University, Taiyuan, 030006, China}

\author{Gerardo Adesso\orcidlink{0000-0001-7136-3755}}
\email{gerardo.adesso@nottingham.ac.uk}
\affiliation{School of Mathematical Sciences and Centre for the Mathematics and Theoretical Physics of Quantum Non-Equilibrium Systems, University of Nottingham, University Park, Nottingham NG7 2RD, United Kingdom}
\date{\today}

\begin{abstract}
Here we propose a general protocol for dual-channel lock-in detection of high-frequency ac signals. We find that the effect of a high-frequency target signal can be modulated through the application of rapidly oscillating driving fields. Based on this mechanism, we develop a quantum dual-channel lock-in detection protocol for high-frequency signals, which not only extends the accessible frequency range of quantum sensing but also enables the simultaneous estimation of the signal amplitude and initial phase. Furthermore, we present a feasible implementation scheme of the protocol based on nitrogen-vacancy centers in diamond. Numerical simulations demonstrate that the proposed protocol can effectively filter out background noise and significantly improve the signal-to-noise ratio. Our results provide a promising approach for realizing noise-resistant detection of weak signals in the high-frequency regime.
\end{abstract}
\maketitle

\section{Introduction}
In recent years, quantum sensing \cite{RMP2017} has attracted considerable attention owing to its potential to achieve unprecedented sensitivity, precision, and spatial resolution. A variety of quantum sensing protocols \cite{Taylor08,Saywell23,Cappellaro22,Zeng2024,Liao2025,Chaudhry2025,Louzon2025,ZHANGLJ2024,Mihailescu25,Lvovsky2026} have been proposed for measuring physical quantities such as magnetic fields, electric fields, temperature, and acceleration. Among the diverse applications of quantum sensing, the detection of ac signals is of fundamental importance for a wide range of applications, such as magnetic resonance spectroscopy \cite{Aslam17,Munuera17,Boss2017,DU2024,Schmitt17,Glenn18,DJF2024}, electromagnetic-field characterization \cite{Wang22,Meinel21,RB20}, and weak-force sensing \cite{Shaniv2017}.

In quantum sensing, conventional approaches for detecting ac signals are typically based on spin-echo or dynamical-decoupling sequences \cite{Taylor08,Nature2011,Genov20,Lange2011,Kotler13}. These approaches operate by measuring the dynamical phase accumulated by the state of a two-level quantum probe exposed to the target field, where periodic pulse operations are applied to flip the probe state and enable continuous phase accumulation despite the oscillatory nature of ac signals. When the initial phase of the target signal is known, it is straightforward to adjust the pulse sequence such that the pulse operations coincide with the nodes of the target signal, thereby ensuring deterministic phase accumulation. However, when the initial phase of the target signal is unknown, the measurement outcome depends on the unknown initial phase. This indicates that the detection process must be repeated many times, and the amplitude of the target signal can only be extracted through statistical analysis of the measurement results \cite{RMP2017,Kotler13}. Consequently, a large number of measurements are required to obtain reliable estimates, leading to increased time overhead and reduced detection efficiency. 

To address this issue, the principle of the classical dual-channel lock-in amplifier \cite{Meade} has been extended to the quantum regime. Quantum versions of dual-channel lock-in detection protocols have been proposed \cite{PRB13,CP2024,Zhuang2024}, in which the amplitude and initial phase of the target signal can be simultaneously determined using two orthogonal periodic multi-pulse sequences. Although these protocols are highly effective for detecting ac signals in the low-frequency range, it is challenging to apply them in the high-frequency range. The difficulty originates from the inherent limitations of pulse-based detection protocols. In practice, the Rabi frequency is always finite, implying that control pulses possess a non-negligible duration. When the frequency of the target signal becomes sufficiently high, the pulse duration may become comparable to the signal period. Under such conditions, nonideal effects arising from finite pulse duration can significantly degrade the sensing performance \cite{PRX2015,Ezzell23}. Therefore, developing a quantum dual-channel lock-in detection protocol that operates in the high-frequency regime remains an open issue in quantum sensing.

In this paper, we propose a general protocol for the dual-channel lock-in detection of high-frequency ac signals by employing rapidly oscillating driving fields. We investigate the dynamics of a quantum probe subjected to a high-frequency target signal and a rapidly oscillating driving field operating at the same frequency. By using the Magnus expansion \cite{mg1,mg2,mg3}, we derive an effective Hamiltonian whose strength is proportional to the product of the target-signal amplitude and the driving-field strength, and is further modulated by their relative phase. Based on this result, we develop a high-frequency dual-channel lock-in detection protocol. The key advantage of our protocol is that it enables lock-in detection at target-signal frequencies far exceeding the maximum achievable Rabi frequency of the quantum probe, thereby offering a promising route to extending the operational frequency range of a wide variety of quantum sensing platforms. Moreover, our protocol enables the direct extraction of both the amplitude and the initial phase of the target signal from only two population measurements, thereby eliminating the need for statistical analysis of a large number of measurement outcomes and significantly improving the detection efficiency. As a concrete example, we present an implementation scheme of high-frequency dual-channel lock-in detection based on nitrogen-vacancy (NV) centers in diamond \cite{Doherty13}. We simulate the performance of the proposed protocol under the influence of different types of background noise. Numerical results demonstrate that our protocol can significantly improve the signal-to-noise ratio (SNR).

\section{THE GENERAL THEORY} \label{GT}
We consider a two-level quantum probe interacting with a driving field and a target ac signal. Under the rotating-wave approximation, the Hamiltonian of the two-level probe takes the form

\begin{align}\label{eq1}
H(t)=H_{C}(t)+H_{S}(t)=\frac{\Omega(t)}{2}\sigma_{x}+\Delta(t)\sigma_{z},
\end{align}
where $\sigma_{x}$ and $\sigma_{z}$ are standard Pauli operators acting on $\ket{0}$ and $\ket{1}$, $\Omega(t)$ is the Rabi frequency of the driving field, and $\Delta(t)$ is the detuning induced by the target signal. The target signal is assumed to be a sinusoidal signal with a known frequency in the hundreds-of-megahertz range and unknown amplitude and initial phase. Thus, the detuning in Eq.~(\ref{eq1}) can be expressed as

\begin{align}\label{eq2}
\Delta(t)=A\cos(\omega t+\varphi),
\end{align}
where $\omega$ is the frequency, $\varphi$ is the initial phase, and $A$ is the amplitude of the target signal. Since the frequency of the target signal lies well beyond the operating range of conventional dynamical-decoupling-based sensing protocols, we propose a dual-channel lock-in detection protocol specifically designed for high-frequency signals.

To realize lock-in detection in high-frequency regime, the Rabi frequency of the driving field in Eq.~(\ref{eq1}) is set to a rapidly oscillating trigonometric function with the same frequency as the target signal, which takes the form of
\begin{align}\label{eq3}
\Omega(t)=2B\sin(\omega t+\phi_{r}),
\end{align}
where $\phi_{r}=0$ or $\pi/2$. In this case, the Hamiltonian of the probe can be rewritten as
\begin{align}\label{eq4}
H(t)=B\sin(\omega t+\phi_{r})\sigma_{x}+A\cos(\omega t+\varphi)\sigma_{z}.
\end{align}
The evolution of the probe governed by the above Hamiltonian can be derived using the Magnus expansion \cite{mg1,mg2,mg3}.

According to the Magnus expansion, the time-evolution operator associated with the Hamiltonian $H(t)$ can be written in the form
\begin{equation}\label{Eq5}
U(t)=\exp\left[\sum_{n=1}^{\infty}M_n(t)\right].
\end{equation}
The first two terms of the Magnus expansion are given by
\begin{align}\label{Eq6}
&M_{1}(t)=-i\int^{t}_{0}H(t_{1})dt_{1},
\notag\\
&M_{2}(t)=-\frac{1}{2}\int^{t}_{0}dt_{1}\int^{t_{1}}_{0}dt_{2}[H(t_{1}),H(t_{2})].
\end{align}
For a sufficiently short evolution time, the evolution is dominated by the first few terms of the Magnus expansion. Here, we calculate the first two terms of the Magnus expansion over one period of the target signal, i.e., $T=2\pi/\omega$. Substituting Eq.~(\ref{eq4}) into Eq.~(\ref{Eq6}), we can obtain
\begin{align}\label{Eq7}
M_{1}(T)=0,M_{2}(T)=i AB \frac{T^{2}}{2\pi}\cos(\phi_{r}-\varphi)\sigma_{y}.
\end{align}
For the Hamiltonian in Eq.~(\ref{eq4}), the higher-order Magnus terms satisfy $M_n(T)=\mathcal{O}(T^n)$. Therefore, these terms become negligible when detecting high-frequency target signals.

\begin{figure*}[th]
	\includegraphics[width=0.95\textwidth]{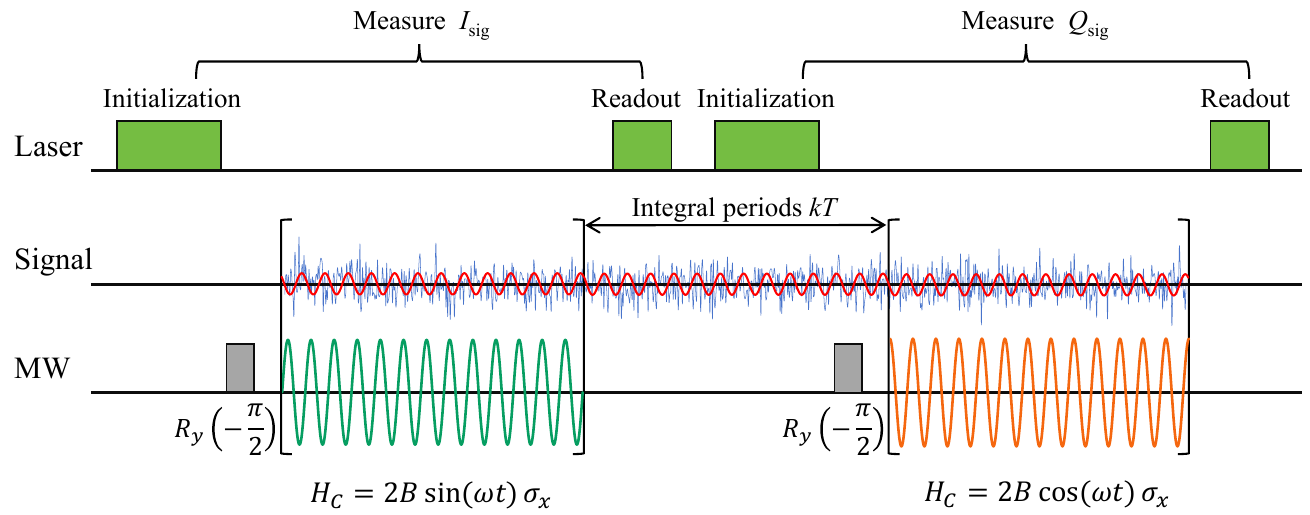}
	\caption{Schematic diagram of the high-frequency quantum dual-channel lock-in detection protocol based on NV centers. The top row shows the laser pulse sequence used to initialize and read out the electron spin state of NV centers. The middle row shows the target signal and background noise. The bottom row shows the driving microwave field used to implement the detection process and gate operations.}\label{Fig1}
\end{figure*}

By using Eqs.~(\ref{Eq5}) and (\ref{Eq7}), the effective Hamiltonian over one period of the target signal can be obtained as
\begin{align}\label{Eq8}
H_{\mathrm{eff}}=-\frac{AB}{\omega}\cos(\phi_{r}-\varphi)\sigma_{y}.
\end{align}
Thus, the evolution of the two-level probe under the influence of both the driving field and the target signal can be effectively described by a static effective Hamiltonian whose strength is proportional to the product of the amplitudes of the driving field and the target signal and is tunable via the relative phase $\phi_{r} - \varphi$. This is a crucial result for achieving lock-in detection in the high-frequency regime.

Based on the above results, we present the details of the dual-channel lock-in detection protocol using rapidly oscillating driving. Under the effective Hamiltonian in Eq.~(\ref{Eq8}), the two-level quantum probe evolves for an integer number of target-signal periods, $\tau=NT$, undergoing a rotation about the $y$ axis on the Bloch sphere. The rotation angle takes the form
\begin{align}\label{Eq9}
\theta(A,\varphi)=-2AB \frac{NT^{2}}{2\pi}\cos(\phi_{r}-\varphi),
\end{align}
which depends on the parameters to be measured, i.e., the amplitude and initial phase of the target signal. To realize lock-in detection, the quantum probe is prepared in the state
$\ket{\psi(0)}=\frac{1}{\sqrt{2}}(\ket{0}+\ket{1})$
and evolves under the Hamiltonian in Eq.~(\ref{eq4}) for an integer number of signal periods, $\tau=NT$, where $N$ is a sufficiently large positive integer. The final state of the quantum probe can be obtained as
\begin{align}\label{Eq10}
\ket{\psi(\tau)}=\frac{1}{\sqrt{2}}\left[\left(\cos\frac{\theta}{2}-\sin\frac{\theta}{2}\right)\ket{0}+\left(\cos\frac{\theta}{2}+\sin\frac{\theta}{2}\right)\ket{1}\right].
\end{align}
The populations of states $\ket{0}$ and $\ket{1}$ are given by
\begin{align}\label{Eq11}
P_{0}=\frac{1-\sin\theta}{2},~P_{1}=\frac{1+\sin\theta}{2}.
\end{align}
Thus, the rotation angle $\theta$ can be determined from the state populations. If the phase of the Rabi frequency of the driving field is set to $\phi_{r}=0$, the rotation angle is given by
\begin{align}\label{Eq12}
\theta_{\phi_{r}=0}(A,\varphi)=-2AB \frac{NT^{2}}{2\pi}\cos\varphi.
\end{align}
Thus, one can determine the value of
\begin{align}\label{Eq13}
I_{\mathrm{sig}}\equiv A\cos\varphi=-\frac{\pi\theta_{\phi_{r}=0}}{NBT^{2}},
\end{align}
by measuring the populations of states $\ket{0}$ and $\ket{1}$. Similarly, if the phase of the Rabi frequency of the driving field is set to $\phi_{r}=\pi/2$, the rotation angle is given by
\begin{align}\label{Eq14}
\theta_{\phi_{r}=\pi/2}(A,\varphi)=-2AB \frac{NT^{2}}{2\pi}\sin\varphi.
\end{align}
The value of
\begin{align}\label{Eq15}
Q_{\mathrm{sig}}\equiv A\sin\varphi=-\frac{\pi\theta_{\phi_{r}=\pi/2}}{NBT^{2}},
\end{align}
can also be determined by measuring the state populations. By using Eqs.~(\ref{Eq13}) and (\ref{Eq15}), the amplitude and initial phase of the target signal can be obtain as
\begin{align}\label{Eq16}
A=\sqrt{I_{\mathrm{sig}}^{2}+Q_{\mathrm{sig}}^{2}},~~\varphi=\arctan\left(\frac{Q_{\mathrm{sig}} }{I_{\mathrm{sig}}}\right).
\end{align}
Therefore, quantum dual-channel lock-in detection of high-frequency ac signals can be realized by implementing the evolution processes described above and measuring the population of probe states.

\section{THE IMPLEMENTATION SCHEME}
While the general framework of high-frequency lock-in detection has been established in the previous section, we now illustrate its feasibility and application through a concrete implementation scheme. Nitrogen-vacancy (NV) centers in diamond \cite{Doherty13} have emerged as powerful magnetic-field sensors owing to their fast coherent spin manipulation, efficient optical initialization and readout, and ability to operate at room temperature. Conventional dynamical-decoupling-based ac field sensing protocols using NV centers are typically confined to the frequency range from kilohertz to several megahertz. In contrast, our protocol extends the accessible sensing range to hundreds of megahertz.

Here, we consider the electron spin of a single negatively charged NV center in an n-type $^{12}$C-enriched diamond \cite{Herbschleb19} as the quantum probe, and assume that the host $^{15}\mathrm{N}$ nuclear spin is polarized \cite{ref10}. In this case, the system has a spin-triplet ground state with sublevels $\ket{m_{s}=0}$ and $\ket{m_{s}=\pm1}$. The Hamiltonian of the negatively charged NV center in a static magnetic field $B_{0}$ applied along the NV axis is given by $H_{\mathrm{NV}}=DS_{z}^{2}+\gamma_{e}B_{0}S_{z}$, where $D=2\pi\times2.87~\mathrm{GHz}$ is the zero-field splitting, $\gamma_{e}=2\pi\times28~\mathrm{GHz/T}$ is the gyromagnetic ratio of the electron spin, and $S_{z}$ denotes the $z$-component of the spin-1 operator of the NV center electron spin in the basis $\{\ket{m_{s}=+1},\ket{m_{s}=0},\ket{m_{s}=-1}\}$. Thus, the degeneracy between the states $\ket{m_{s} = \pm 1}$ can be lifted via Zeeman splitting of $2\gamma_{e} B_{0}$, induced by the static magnetic field $B_{0} \approx 500~\text{G}$.

To realize lock-in detection, we take the two Zeeman levels $\ket{m_{s}=-1}\equiv\ket{0}$ and $\ket{m_{s}=0}\equiv\ket{1}$ as the basis states of the quantum probe, and restrict the dynamics to this subspace. In this subspace, the Hamiltonian reduces to $H_{NV}=\omega_{0}\sigma_{z}/2$, with $\omega_{0}=D-\gamma_{e}B_{0}\approx2\pi\times1.47~\text{GHz}$. By employing optical pumping cycles \cite{Robledo11}, the electron spin can be initialized into the $\ket{m_s=0}\equiv\ket{1}$ state, and the spin-state populations can be read out via the detection of spin-dependent fluorescence emission.

The control operations for preparing the initial state and implementing the control Hamiltonian $H_{\mathrm{C}}(t)$ can be realized by applying a resonant microwave driving field with frequency $\omega_0$ that couples the two Zeeman sublevels $\ket{m_s = -1}$ and $\ket{m_s = 0}$. Due to the Zeeman effect, the energy levels $\ket{m_{s}=\pm1}$ shift linearly with the magnetic field along the NV axis. Thus, the magnetic field $B_{m}(t)$ sensed by the NV center induces a detuning described by the signal Hamiltonian $H_{\mathrm{S}}(t)=\Delta(t)\sigma_{z}$, where $\Delta(t)=-\gamma_{e} B_{m}(t)/2$.

With these concepts, we can present the workflow of the high-frequency dual-channel lock-in detection scheme. As shown in Fig.~\ref{Fig1}, the implementation scheme of high-frequency dual-channel lock-in detection using a single NV center consists of two stages. In the first stage, the in-phase component $I_{\mathrm{sig}}$ is measured, while in the second stage, the quadrature component $Q_{\mathrm{sig}}$ is measured. In the first stage, the electron spin is first initialized to the state $\ket{1}$ and subsequently prepared in the state $\frac{1}{\sqrt{2}}(\ket{0}+\ket{1})$ by a $R_{y}(-\pi/2)$ gate. Then, the state of electron spin evolves under the combined influence of the target magnetic signal and the microwave driving field with $\phi_{r}=0$ for an integer number of periods of the target signal. Finally, the in-phase component $I_{\mathrm{sig}}$ can be extracted from measurements of the spin-state populations. In the second stage, the state-preparation and readout procedures are identical to those in the first stage, except that the phase of Rabi frequency of the microwave driving field is set to $\phi_r=\pi/2$. The start of the second-stage driving field is delayed from the end of the first-stage driving field by an integer multiple of the target-signal period. During this delay, the readout of the first stage and the state preparation of the second stage are carried out. The quadrature component $Q_{\mathrm{sig}}$ can be extracted from measurements of the spin-state populations at the end of the second stage. Here, we focus on the detection scheme based on a single NV center. In principle, if one can address, control, and read out two spatially close NV centers independently, $I_{\mathrm{sig}}$ and $Q_{\mathrm{sig}}$ can be measured synchronously. 
\begin{figure}[tb]
  \includegraphics[scale=0.3]{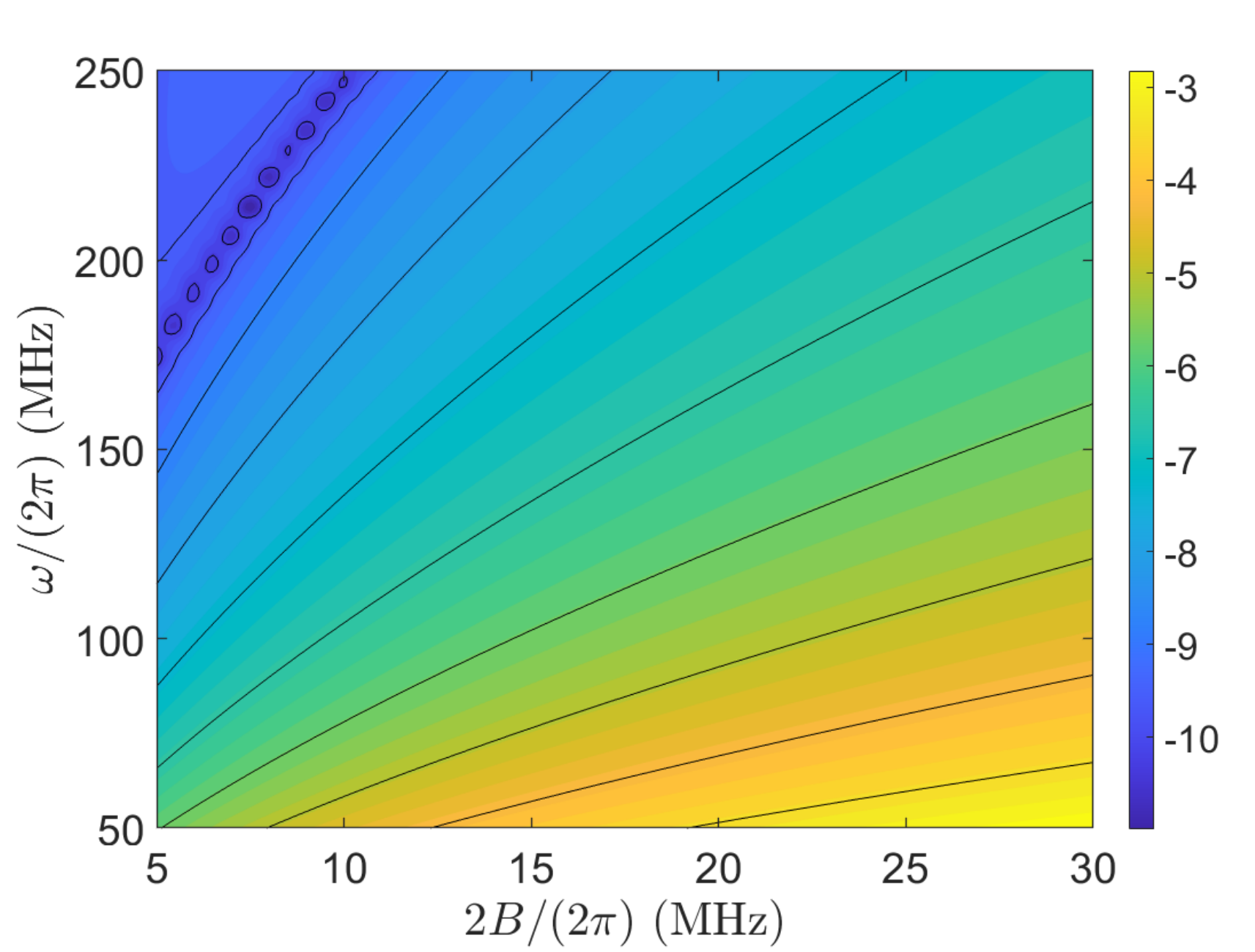}
    \caption{The validity of the second-order Magnus approximation. The quantity $\log_{10}(1-F)$ is plotted as a function of signal frequency $\omega$ and driving-field amplitude $2B$.}
   \label{Fig2}
\end{figure}

\section{DISCUSSION}
In this section, we evaluate the performance of the proposed high-frequency dual-channel lock-in detection protocol. First, we numerically verify the validity of the second-order Magnus approximation underlying the theoretical framework of the protocol. We then investigate the performance of the proposed lock-in detection in the presence of different types of background noise.

To quantify the validity of the second-order Magnus approximation, we calculate the infidelity, $1-F$, where the fidelity $F$ between the actual and approximate evolution operators is defined as
\begin{equation}
F=\frac{1}{2}\left|\text{Tr}[U_{\text{eff}}(\tau)U^{\dag}_{A}(\tau)]\right|,
     \label{eq}
\end{equation}
in which $U_{\text{eff}}(\tau)=\exp[-iH_{\text{eff}}\tau]$ is the effective evolution operator derived from the second-order Magnus approximation, while $U_{A}(\tau)$ denotes the actual evolution operator obtained by numerically solving the time-dependent Schr\"{o}dinger equation governed by the Hamiltonian in Eq.~(\ref{eq4}). Because the signal amplitude $A$ is a small quantity, the validity of the Magnus approximation is primarily determined by the frequency $\omega$ and the driving-field amplitude $2B$. Therefore, we scan the frequency $\omega$ and the driving-field amplitude $2B$ over a typical experimentally achievable range and plot $\log_{10}(1-F)$ as a function of these two parameters. In the simulations, the signal amplitude is set to $A = 10\ \text{kHz}$, the phases of driving-field Rabi frequency and target signal are set to $\phi_{r}=\varphi=0$, and the evolution time is set to $2000$ periods of the target signal, i.e., $\tau=N(2\pi/\omega)$ with $N=2000$. As shown in Fig.~\ref{Fig2}, the infidelity $1-F$ between the actual and approximate evolution operators remains below $10^{-3}$ over most of the parameter range considered here, indicating that the second-order Magnus approximation is valid over the typical experimentally achievable parameter range.

We next discuss the performance of the proposed protocol in the presence of background noise. In practice, weak signals are often obscured by background noise, making their reliable detection and characterization a major challenge. For example, in NV-center-based microscale and nanoscale magnetic resonance spectroscopy \cite{Aslam17,Munuera17,DJF2024,DU2024}, the target signal is typically the nuclear magnetic resonance signal generated by a cluster of nuclear spins outside the diamond lattice. Such signals are usually buried in magnetic noise originating from the diamond surface, the surrounding nuclear-spin bath, and electronic equipment in the experimental environment. Because the target signal is often extremely weak, the ability to extract it from the noisy background is of central importance. 

To quantitatively assess the measurement error induced by background noise, we calculate the root mean squared errors (RMSEs) of the amplitude and phase, which are defined as
\begin{align}
\mathrm{RMSE}_{A}=\sqrt{\langle(\tilde{A}-A)^{2}\rangle},~\mathrm{RMSE}_{\varphi}=\sqrt{\langle(\tilde{\varphi}-\varphi)^{2}\rangle},
\label{RMSE}
\end{align}
in which $A$ and $\varphi$ denote the true amplitude and phase, respectively, while $\tilde{A}$ and $\tilde{\varphi}$ denote the corresponding measured values, and $\langle\cdot\rangle$ denotes the statistical average over multiple measurements. To evaluate the noise-suppression capability of the proposed lock-in detection protocol, we define the input and output SNRs and calculate the signal-to-noise improvement ratio (SNIR). The input SNR of the signal sensed by the quantum probe is defined as
\begin{align}
\begin{split}
\text{SNR}_{\mathrm{in}}=10\log_{10}(P_{S}/P_{N}),
\end{split}
\label{SNR1}
\end{align}
in units of decibel (dB), where $P_{S}$ and $P_{N}$ denote the powers of the target signal and background noise, respectively. Similarly, the output SNR of the lock-in detection process is defined as
\begin{align}
\begin{split}
\text{SNR}_{\mathrm{out}}=10\log_{10}(A^{2}/\langle(\tilde{A}-A)^{2}\rangle).
\end{split}
\label{SNR2}
\end{align}
The signal-to-noise improvement ratio (SNIR) can be further defined as
\begin{align}
\begin{split}
\mathrm{SNIR}=\text{SNR}_{\mathrm{out}}-\text{SNR}_{\mathrm{in}}.
\end{split}
\label{SNIR}
\end{align}
A larger SNIR indicates that the detection protocol is more effective in extracting information about the target signal from background noise.
\begin{figure}[tb]
  \includegraphics[scale=0.56]{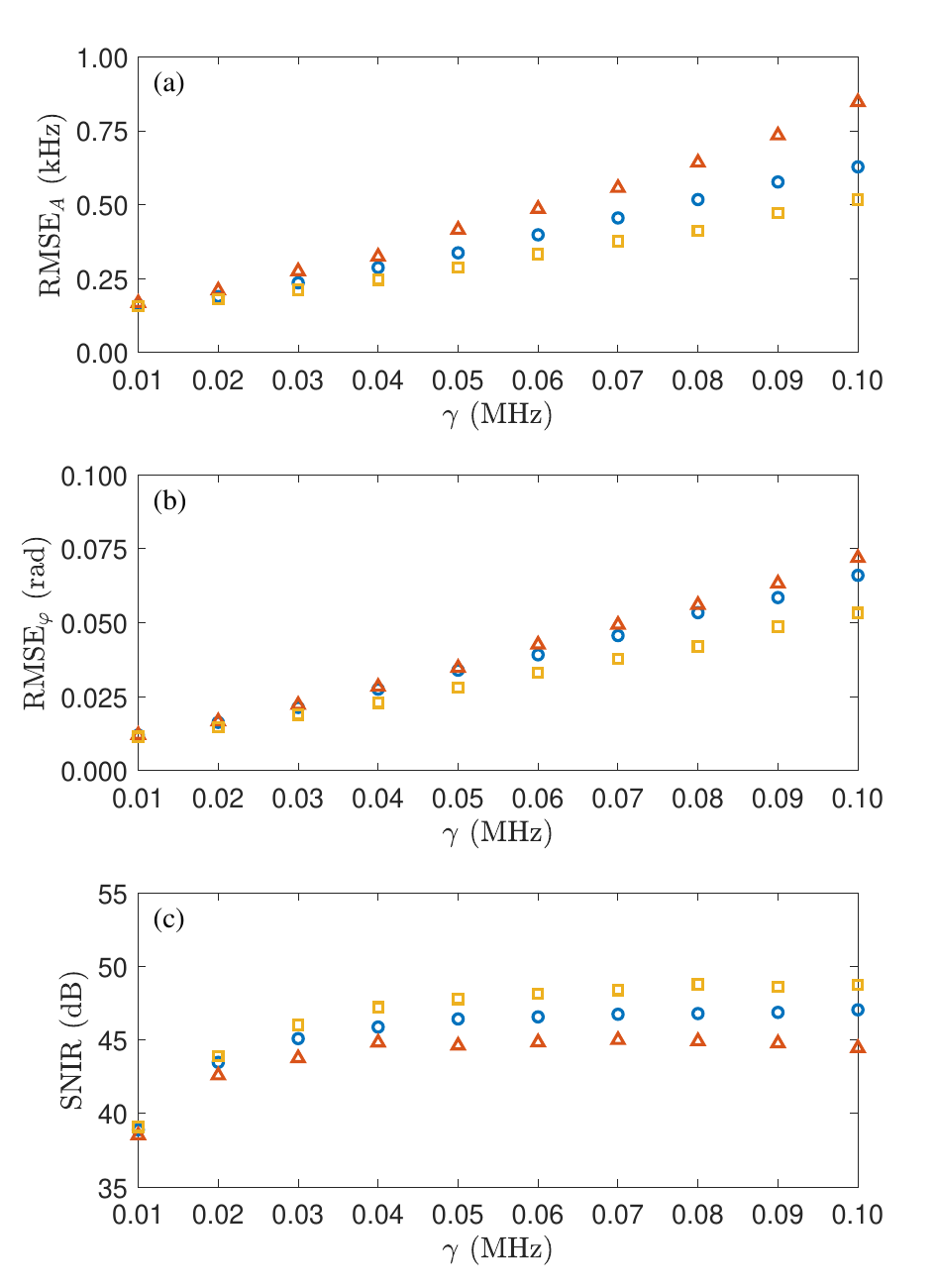}
    \caption{Performance of the proposed high-frequency dual-channel lock-in detection protocol under different types of background noise. (a) RMSE of the amplitude. (b) RMSE of the phase. (c) SNIR of the proposed protocol. White noise (blue circles), $1/f$ noise (red triangles), and harmonic noise (orange squares).}
   \label{Fig3}
\end{figure}

As representative background noise sources, we consider white noise, $1/f$ noise, and harmonic noise, and numerically investigate the performance of the proposed protocol. White noise is characterized by a constant power spectral density over the entire frequency range and is one of the most common background noise sources in weak-signal detection \cite{wnoise}. $1/f$ noise is a ubiquitous low-frequency noise source that often dominates the noise spectrum at low frequencies \cite{fnoise}. Harmonic noise commonly arises from electrical equipment and power systems and is characterized by discrete spectral components at the fundamental frequency and its harmonics \cite{hnoise}. We assume that the target signal and the background noise share the same physical nature and couple to the probe through the same interaction mechanism. Thus, the background noise can be taken into account by introducing an additional noise term $H_{N}=\Delta^{\prime}(t)\sigma_{z}$ into the Hamiltonian in Eq.~(\ref{eq1}). In the simulations, the target signal is assigned an amplitude of $A = 10\ \text{kHz}$, an initial phase of $\varphi=\pi/4$, and a frequency of $\omega = 2\pi \times 100\ \text{MHz}$. The control Hamiltonian $H_{\text{C}}(t)$ is set according to the theory in Section \ref{GT}. The peak value of the Rabi frequency of the driving field is set to $2B = 2\pi \times 20~\text{MHz}$. Each measurement of $I_{\mathrm{sig}}$ or $Q_{\mathrm{sig}}$ involves an evolution time of $\tau=\SI{20}{\micro\second}$, corresponding to $N=2000$ periods of the target signal. We numerically investigate the performance of the proposed high-frequency lock-in detection protocol under different noise models and noise intensities. The RMSEs and SNIR are evaluated from 2500 independent numerical realizations of the detection process.

In the simulations, the RMSEs are calculated using Eq.~(\ref{RMSE}), and the SNIR is calculated using Eq.~(\ref{SNIR}). The power of the target signal is $P_{S}=A^{2}/2$. The white noise is generated from a normal distribution with zero mean and standard deviation $\gamma$, and the power of the white noise is $P_{N}=\gamma^{2}$. The $1/f$ noise is generated using the Voss-McCartney algorithm \cite{fnoise1} with zero mean and standard deviation $\gamma$, and the power of the $1/f$ noise is $P_{N}=\gamma^{2}$. For harmonic noise, the noise Hamiltonian can be expressed as
\begin{align}
\begin{split}
H_{N}=\sum^{\infty}_{n=2}N_{n}\cos(n\omega t+\varphi_{n})\sigma_{z},
\end{split}
     \label{hnoise}
\end{align}
where $N_{n}$ and $\varphi_{n}$ are the amplitudes and phases of the $n$-th harmonic. In this case, the power of the harmonic noise is $P_{N}=(\sum_{n}N_{n}^{2})/2$. Here, only the second and third harmonic components are taken into account, with $N_{2}=N_{3}=\gamma$. The phases $\varphi_{2}$ and $\varphi_{3}$ are randomly assigned in each realization of simulation.

The performance of the proposed high-frequency dual-channel lock-in detection protocol in the presence of different types of background noise is illustrated in Fig.~\ref{Fig3}. As shown in Figs.~\ref{Fig3}(a) and \ref{Fig3}(b), the RMSEs of both the amplitude and phase increase with the noise intensity, as expected. However, as shown in Fig.~\ref{Fig3}(c), the SNIR also increases with increasing noise intensity in the low-noise-intensity regime. This phenomenon can be attributed to the relatively high input SNR in the low-noise-intensity regime, where the room for SNR improvement is limited. As the noise intensity increases, the advantages of the proposed protocol become more evident. In the high-noise-intensity regime, the SNR is improved significantly even when the noise power is an order of magnitude larger than the signal power. This indicates that the proposed protocol can effectively extract the target-signal information in the presence of strong background noise.

The numerical results presented in this section demonstrate that the proposed high-frequency dual-channel lock-in detection protocol is valid over the typical experimentally achievable parameter range. Furthermore, it can effectively extract the target-signal information from a noisy background and significantly improve the SNR, even when the background noise is much stronger than the target signal.

\section{CONCLUSION}
In conclusion, we have proposed a general protocol for the dual-channel lock-in detection of high-frequency ac signals, which enables the simultaneous measurement of the amplitude and phase of the target signal. In this protocol, signal modulation is achieved by applying rapidly oscillating driving fields to the quantum probe, enabling the accumulation of the signal-induced effect without relying on pulse operations. Thus, our protocol enables the lock-in detection of high-frequency ac signals with frequencies far exceeding the maximum achievable Rabi frequency of the quantum probe, thereby extending the accessible frequency range across a wide variety of quantum sensing platforms, such as Rydberg atoms \cite{ZHANGLJ2024}, trapped ions \cite{TI21}, and superconducting qubits \cite{Ezzell23}. As a typical example, we further present a practical implementation of our protocol based on NV centers in diamond, which demonstrates the feasibility of the protocol under current experimental conditions. Simulation results demonstrate that our protocol exhibits strong robustness against different types of background noise and achieves a significant improvement in the SNR. These features make it a promising tool for high-frequency quantum sensing in noisy environments. The ability to detect weak high-frequency ac signals in noisy environments opens promising prospects for applications such as nanoscale spectroscopy \cite{Aslam17,Munuera17}, microwave-field characterization \cite{Wang22,Meinel21}, and quantum noise spectroscopy \cite{Bylander11,Wudarski23}. Furthermore, it would be interesting to extend the high-frequency dual-channel lock-in detection protocol to multi-level probes \cite{ML17,ML26} and many-body quantum systems \cite{PR2025}, where additional sensing capabilities and functionalities may emerge.

\medskip

\begin{acknowledgments}
\noindent\begin{minipage}{\columnwidth}
\hspace*{1em}
K.Z.L. gratefully acknowledges support from the National Natural Science Foundation of China (Grant No.~62405210) and the Fundamental Research Program of Shanxi Province (Grant No.~202403021212252). L.T.X. gratefully acknowledges support from the National Natural Science Foundation of China (Grants No.~62127817 and No.~U23A20380) and the Science and Technology Major Special Project of Shanxi Province (Grant No.~202201010101005). G.A. gratefully acknowledges support from the Engineering and Physical Sciences Research Council (EPSRC Grant No.~EP/X010929/1).
\end{minipage}
\medskip
\end{acknowledgments}

\end{document}